\documentclass[aps,prd,nofootinbib,twocolumn,showpacs,a4paper,floatfix]{revtex4}
\usepackage{epsfig}
\usepackage{amssymb}
\usepackage{amsmath}
\usepackage{amsfonts}
\usepackage{graphicx}
\def\dmtwon{\Delta m^2_{21}}
\def\dmthon{\Delta m^2_{31}}
\def\dmthtw{\Delta m^2_{32}}
\def\dmsol{\Delta m_\odot^2}

\def\dmatm{\Delta m_{\rm atm}^2}
\def\thetasol{\theta_\odot}
\def\thetaatm{\theta_{\rm atm}}
\def\evolt{{\rm eV}}
\def\eV2{{\rm eV^2}}

\begin{document}

\title{Neutrino Masses, Mixing and New Physics Effects}
\author{J. A. Aguilar--Saavedra, G. C. Branco, F. R. Joaquim}
\affiliation{Departamento de F\'{\i}sica, and GTFP,
Instituto Superior T\'ecnico,
P-1049-001 Lisboa, Portugal}
\date{\today}

\begin{abstract}
We introduce a parametrization of the effects of radiative corrections
from new physics on the charged lepton and neutrino mass matrices, studying
how several relevant quantities describing the pattern of neutrino masses and
mixing are affected by these corrections. We find that the ratio
$\omega \equiv \sin \theta/\tan \theta_\mathrm{atm}$ is 
remarkably stable, even when relatively large corrections are added to the
original mass
matrices. It is also found that if the lightest neutrino has a mass around 0.3
eV, the pattern of masses and mixings is considerably more stable under
perturbations than for a lighter or heavier spectrum.
We explore the consequences of perturbations on some flavor relations given
in the literature. In addition, for a quasi-degenerate neutrino
spectrum it is shown 
that: (i) starting from a bi-maximal mixing scenario, the corrections to the
mass matrices keep $\tan \theta_\mathrm{atm}$ very close to unity while they
can lower
$\tan \theta_\odot$ to its measured value; (ii) beginning from a scenario with
a vanishing Dirac phase, corrections can induce a Dirac phase large
enough to yield $CP$ violation observable in neutrino oscillations.
\end{abstract}

\pacs{14.60.Pq,12.60.-i,11.30.Hv,11.30.Er}


\maketitle

\section{Introduction}
\label{sec:1}

Our present knowledge of neutrino masses and mixing is mainly
provided by various neutrino oscillation experiments, which give us information
on the two independent mass squared differences, as well as on the
three angles characterizing the leptonic mixing matrix. In the future, the study
of $CP$ violation in neutrino oscillations may allow to determine
the Dirac-type phase entering in the leptonic mixing matrix, and neutrinoless
double $\beta$
decay experiments may provide the value of the effective Majorana mass.
Despite the great achievements of oscillation
experiments, there is still much to be learned about neutrinos.
One of the most problematic issues in neutrino physics is the lack
of information on the mass spectrum, since the mass squared differences do not
fix the absolute scale of neutrino masses. Indeed, the spectrum can exhibit a
strong hierarchy, as in the case of quarks and charged leptons, or on the
contrary, be quasi-degenerate.

At present, there are different extensions of the standard model (SM) which
propose mechanisms for the generation of neutrino masses through the
enlargement of the SM particle content. The addition of heavy right-handed
neutrino singlets constitutes a simple and economical way to give left-handed
neutrinos small masses through the seesaw mechanism
\cite{seesaw1,seesaw2,seesaw3,seesaw4}. Another simple possibility relies on the
extension of the scalar sector with a heavy $\mathrm{SU}(2)_L$ triplet, with or
without supersymmetry (SUSY)
\cite{Cheng:qt,Gelmini:1980re,Wetterich:1981bx,Rossi:2002zb}.
Within supersymmetric models, neutrino masses can also arise
from $R$-parity violating interactions \cite{Diaz:2003as}, where the atmospheric
and solar neutrino mass scales are generated at the tree level and radiatively,
respectively. Recently, a new supersymmetric source of neutrino masses and
mixings has been found considering non-renormalizable lepton number violating
interactions in the K{\"a}hler potential \cite{Casas:2002sn} rather than in the
effective superpotential. All the above scenarios predict the existence of light
Majorana massive neutrinos. Suppressed Dirac neutrino masses are unnatural in
conventional theories since they usually require extremely small Yukawa
couplings. However, this problem can be surmounted by endowing the particle
content of the theory with extra fermion or scalar fields and/or introducing
new symmetries \cite{dirac1,dirac1b,dirac2,dirac3,dirac4}. Naturally small Dirac
neutrino masses also arise
in extra-dimensional theories as a consequence of the small overlap between the
wavefunctions of the usual left-handed neutrinos in the brane and the sterile
right-handed ones in the bulk (or in other branes)
\cite{extra1,extra2,extra3,barenboim,extra4,extra5}.

In addition to the mechanism for the generation of neutrino masses,
one of the most intriguing aspects of leptonic physics is the experimental
evidence that two of the lepton mixing angles are large, in contrast to
the small
mixing observed in the quark sector. The deep understanding of the neutrino mass
suppression mechanism and the bi-large leptonic mixing constitutes one of the
most challenging questions in particle physics. A theory of leptonic flavor
should provide a plausible explanation for the bi-large mixing, as well as for
the neutrino mass spectrum. At the same time, it should predict relations
among these quantities. There have been in the literature a large number of
suggestions in this direction, consisting in the introduction of flavor
symmetries or the assumption of specific textures for the leptonic mass
matrices \cite{models,reviews}.

The tree-level predictions of any theory of leptonic flavor are subject
to higher order corrections, which are computable if the particle content and
parameters of the new physics model are specified. Still, in the absence of a
``standard theory'' of lepton flavor, it is pertinent to ask ourselves about
the possible consequences of the (unknown) radiative corrections on the
tree-level predictions of the model. The aim of this paper is to investigate
these effects, focusing on: (i) the modification of the tree-level values for
the ratio of mass squared differences, mixing angles and CP violation
parameters; (ii) the effect on various flavor relations. This task will
be carried out in
two steps. We will first propose a parametrization for the unknown radiative
corrections to the lepton mass matrices, based on rather general arguments on
weak-basis independence. In this parametrization, the formal structure of the
corrections is fixed, up to undetermined complex coefficients. Then, the
possible effects of the radiative corrections is investigated taking these
coefficients as random, and performing a statistical analysis of the average
behavior of the parameters and flavor relations under
consideration.

It is important to note here the difference between our framework and other 
studies in the literature regarding random neutrino matrices
\cite{murayama,vissani,masina,espinosa}. In these
analyses, the neutrino mass matrices are taken as random at the tree level, with
a subsequent discussion of the predictions for the neutrino mass spectrum and
mixing angles. In contrast with this approach, in this paper we take the
tree-level charged lepton and neutrino mass matrices as fixed by some theory of
leptonic flavor, and consider random perturbations (from radiative corrections)
to them \cite{anarch}. These corrections are not completely arbitrary: the
different terms of the perturbations must have a determined formal structure
(which is dictated by weak basis independence) but have random coefficients.

This paper is organized as follows. In Section \ref{sec:2} we briefly review the
current status of neutrino oscillation data. The
parametrization of the unknown corrections to the lepton mass matrices is
derived in Section \ref{sec:3}. In Section \ref{sec:4} we present our results,
discussing (i) the effect of perturbations on masses and mixing parameters; (ii)
the influence on some flavor relations obtained for some specific patterns in
the literature; (iii) the consequences of the corrections in some special limits
of interest. Our conclusions are summarized in Section \ref{sec:5}.

\section{Neutrino oscillations: present status}
\label{sec:2}
All presently available neutrino oscillation data can be accommodated within the
framework of three mixed massive neutrinos
\cite{Giunti:2003qt}.\footnote{We will not consider here
the results from the liquid scintillator neutrino detector (LSND).}
The first KamLAND results
\cite{Eguchi:2002dm} select the large mixing angle MSW solution as the
only surviving explanation for the solar neutrino problem. In addition, dominant
solar neutrino conversion based on non-oscillation solutions are now excluded
\cite{valle1,valle2}. Recently, the Sudbury Neutrino Observatory (SNO) has
released the improved measurements of the salt enhanced phase
\cite{Ahmed:2003kj} which, together with all the solar and KamLAND neutrino
data, allow for a better determination of the oscillation parameters. In
particular, the high-$\Delta m^2$ region ($\dmsol > 10^{-4}\,\eV2$) is now only
accepted at the $3\sigma$ level and maximal solar mixing is ruled out by more
than $5\sigma$, rejecting in this way the possibility of bi-maximal leptonic
mixing
\cite{Balantekin,fogli,maltoni,aliani,petcov,creminelli,deHolanda:2003nj}.
Concerning the atmospheric
neutrino sector, the water Cherenkov Super-Kamiokande (SK) \cite{Fukuda:1998mi}
and long-baseline KEK-to-Kamioka (K2K) \cite{Ahn:2002up} experiments indicate
that neutrino flavor conversion due to neutrino oscillations in the $\nu_\mu
\rightarrow \nu_\tau$ channel provide by far the most acceptable and natural
explanation for the observed $\nu_\mu$ disappearance.

Regarding the absolute values of neutrino masses, the situation is not so
satisfactory. At present, the most stringent direct bound on the neutrino mass
is provided by the Mainz \cite{Bonn:jw} and Troitsk \cite{Lobashev:uu}
experiments, which have set a maximum value for $m_{\nu_e}$ of $2.2\,\evolt$.
Neutrinoless
double beta-decay measurements may also be valuable to disentangle the pattern
of neutrino masses, although in this case certain subtleties have to be
considered \cite{Joaquim:2003pn}.

Throughout this paper we adopt the neutrino mass ordering
$m_1 < m_2 < m_3$ in such a way that, for the hierarchical (HI),
inverted-hierarchical (IH) and quasi-degenerate (QD) neutrino mass
spectra one has
\begin{eqnarray}
\label{ordering}%
\text{HI} & \rightarrow & m_1 \ll m_2\,,\,m_3  \,,~
   \dmsol = \dmtwon  \,; \nonumber \\
\text{IH} & \rightarrow & m_1 \ll m_2\,,\,m_3 \,,~
   \dmsol = \dmthtw \,; \nonumber \\
\text{QD} & \rightarrow & m_1 \simeq m_2 \simeq m_3 \,,~
   \dmsol = \dmtwon  \,,\,
\end{eqnarray}
where $\dmsol$ and $\dmatm = \dmthon \gg \dmsol$ are the solar and
atmospheric neutrino mass squared differences, respectively. At
the $1\sigma$ level, the allowed ranges for $\dmsol$ and $\dmatm$
are \cite{deHolanda:2003nj,Fogli:2003th}
\begin{align}
\label{Deltaranges}%
& \dmsol=(6.5-8.5)\times 10^{-5}\,\evolt^2 \,, \nonumber \\
& \dmatm=(2.6\pm 0.4)\times 10^{-3}\,\evolt^2 \,,
\end{align}
with the best-fit values
\begin{align}
\label{Deltabest}%
& \dmsol=7.13\times 10^{-5}\,\evolt^2 \,,\nonumber \\
& \dmatm=2.6\times 10^{-3}\,\evolt^2 \,.
\end{align}
For three light Majorana neutrinos, the leptonic
mixing matrix $U$ can be written as
\begin{align}
 U  =\left(
\begin{array}{ccc}
U_{e1} & U_{e2} & U_{e3}   \\
U_{\mu 1} & U_{\mu 2} & U_{\mu 3} \\
U_{\tau 1} & U_{\tau 2} & U_{\tau 3}
\end{array}
\right)=U_\delta \, \text{diag}(e^{-i\alpha},e^{-i\beta},1) \,,
\label{U}
\end{align}
where $\alpha$ and $\beta$ are Majorana-type phases and $U_\delta$ can
be parametrized in the form
\begin{align}
\label{Udelta}%
U_\delta= \left(\begin{array}{ccc}
c_{3} c_{2} & s_{3} c_{2} & s_{2} e^{-i \delta} \\
-s_{3} c_{1} - c_{3} s_{1} s_{2}  e^{i \delta} &  c_{3} c_{1} -
s_{3} s_{1} s_{2} e^{i \delta}
& s_{1} c_{2} \\
s_{3} s_{1} - c_{3} c_{1} s_{2} e^{i \delta} & -c_{3} s_{1} -
s_{3} c_{1} s_{2} e^{i \delta} & c_{1} c_{2}
\end{array}\right)
\end{align}
with $s_i \equiv \sin \theta_i$, $c_i \equiv \cos \theta_i$
$(i=1,2,3)$ and $\delta$ the Dirac-type $CP$-violating phase. 
For Dirac neutrinos $U$ reduces to $U_\delta$, due to the absence of
Majorana phases. Depending
on the type of neutrino mass spectrum, the solar, atmospheric and
CHOOZ \cite{Apollonio:1999ae} mixing angles ($\thetasol$,
$\thetaatm$ and $\theta$, respectively) can be extracted from $U$
in the following way: for the HI and QD neutrino mass spectra,
\begin{align}
\label{angdefHI}%
&\tan\thetasol=\frac{|U_{e2}|}{|U_{e1}|}=\tan\theta_3\,\,,\,\,
\tan\thetaatm=\frac{|U_{\mu 3}|}{|U_{\tau 3}|}=\tan\theta_1\,,\nonumber \\
&\sin \theta=|U_{e 3}|=\sin\theta_2\,,
\end{align}
and for an IH spectrum,
\begin{align}
\tan\thetasol=\frac{|U_{e3}|}{|U_{e2}|}\,\,,\,\,
\tan\thetaatm=\frac{|U_{\mu 1}|}{|U_{\tau
1}|}\,\,,\,\,\sin\theta=|U_{e 1}|\,.
\end{align}
In this case the expression of $\theta_\odot$, $\theta_\mathrm{atm}$
and $\theta$ in terms of $\theta_{1-3}$ in the parametrization of
Eq.~(\ref{Udelta}) is not simple.
From the global analyses performed
in Refs.~\cite{deHolanda:2003nj} and \cite{Fogli:2003th}, the
solar and atmospheric mixing angles are constrained to lay in the
$1\sigma$ intervals
\begin{align}
\label{tansilatm}%
\tan^2\thetasol=(0.33-0.47)\,\,,\,\,\sin^2 2 \thetaatm=1.00^{+0.00}_{-0.05}\,,
\end{align}
with the best-fit values
\begin{align}
\label{anglesbest}%
\tan^2\thetasol=0.39\,\,,\,\,\sin^2 2 \thetaatm=1.00\,.
\end{align}
For $\sin\theta$ we quote the result from combined
analysis of the solar neutrino, CHOOZ and KamLAND data performed
in Ref.~\cite{Fogli:2002au},
\begin{align}
\label{chooz}%
\sin\theta < 0.18\,,
\end{align}
with a 95\% confidence level (CL).

The upcoming long-baseline neutrino oscillation experiments will face
the challenge of detecting $CP$-violating effects induced by the
Dirac phase $\delta$ \cite{Lindner:2002vt}.
The difference of the $CP$-conjugated neutrino
oscillation probabilities
$P(\nu_e \rightarrow \nu_\mu) - P(\bar{\nu}_e \rightarrow \bar{\nu}_\mu)$
is proportional to the quantity
\begin{align}
{\cal J} & \equiv {\rm Im}\left[\,U_{e1} U_{\mu 2}
U_{e2}^\ast U_{\mu 1}^\ast\,\right] \nonumber \\
&= \frac{1}{8} \sin 2 \theta_{1} \;
\sin 2 \theta_{2} \; \sin 2 \theta_{3} \; \sin \delta\,.
\label{Jgen1}
\end{align}
Present estimates indicate that for $|\mathcal{J}| \gtrsim 10^{-2}$ it will be
possible to observe $CP$ violation effects in these experiments.

\section{Parametrizing the new physics contributions to lepton masses}
\label{sec:3}

\begin{figure}[htb]
\begin{center}
\begin{tabular}{ccc}
\mbox{\epsfig{file=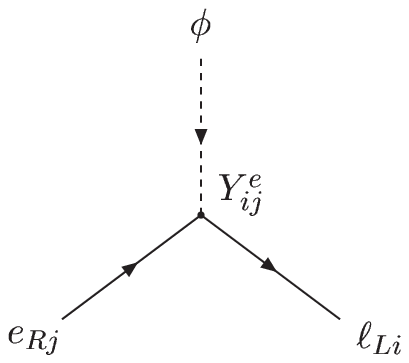,width=3cm,clip=}} & ~~~ &
\mbox{\epsfig{file=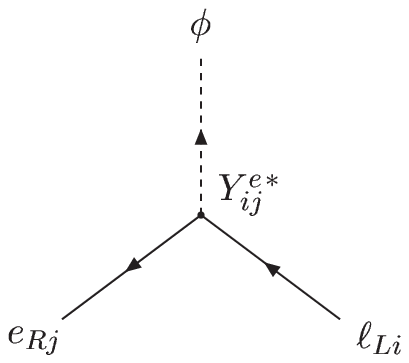,width=3cm,clip=}} \\[0.7cm]
\mbox{\epsfig{file=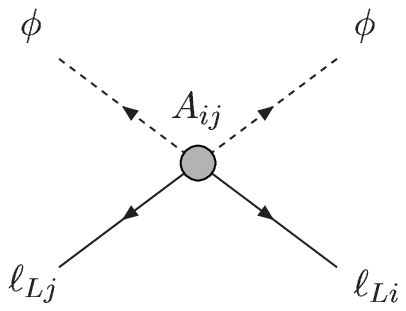,width=3cm,clip=}} & ~~~ &
\mbox{\epsfig{file=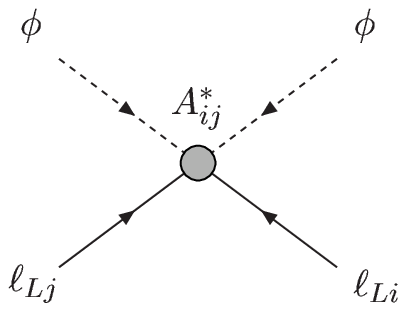,width=3cm,clip=}}
\end{tabular}
\end{center}
\caption{Feynman diagrams corresponding to the vertices in Eq.~(\ref{ec:3.1}).
In each vertex the flavor dependence is explicitly shown.
\label{fig:vertices}}
\end{figure}

Before electroweak symmetry breaking (EWSB), the terms of the Lagrangian 
that originate the charged lepton and light Majorana neutrino masses can be
written as
\begin{equation}
\mathcal{L} = - Y^e_{ij} \, \bar \ell_{Li} \, \phi \, e_{Rj} +
\frac{A_{ij}}{\Lambda} (\bar \ell_{Li} \, i \sigma_2 \, \phi^*) \,
(\phi^\dagger \, i \sigma_2 \, \ell_{Lj}^c) +
\mathrm{H.c.} \,,
\label{ec:3.1}
\end{equation}
where $\ell_{Li}=(\nu_{Li} ~ e_{Li})^T$, $\phi=(\phi^+ ~ \phi^0)^T$ is the
SM Higgs doublet and
$Y^e$ is the usual $3 \times 3$ matrix of the charged lepton Yukawa couplings.
The second term is a neutrino mass operator \cite{weinberg} generated
by physics above the
electroweak scale, being $A$ a $3 \times 3$ symmetric matrix of
dimensionless couplings of order unity and $\Lambda$ the scale at which this
interaction is generated. These vertices can be depicted by the Feynman
diagrams in Fig.~\ref{fig:vertices}, where the flavor dependence of
each vertex is explicitly shown. The arrows in the scalar lines indicate the
flow of the positive charge for the $T=1/2$ component of the doublet.

There are different mechanisms that may lead to the neutrino mass operator
given in Eq.~(\ref{ec:3.1}). In particular, it may
be originated by a heavy scalar triplet $\Delta$, which is conveniently written
in matrix form as
\begin{equation}
\Delta = \left(
\begin{array}{cc} \Delta^+ & \sqrt 2 \Delta^{++} \\
\sqrt 2 \Delta^0 & - \Delta^+
\end{array} \right) \,.
\label{ec:3.2a}
\end{equation}
This scalar triplet couples to the lepton and Higgs doublets as
\begin{equation}
\mathcal{L}_\Delta= Y^\Delta_{ij} \, \bar \ell_i \, \Delta^\dagger \, 
i \sigma_2 \, \ell_j^c + g_{\Delta \phi} \, \phi^\dagger \Delta \, i \sigma_2
\, \phi^* + \mathrm{H.c.} \,,
\label{ec:3.2}
\end{equation}
where $Y^\Delta$ is a matrix of Yukawa couplings and $g_{\Delta \phi}$ has
dimension of mass.
The exchange of the heavy triplet results in an effective neutrino mass
operator, given by
\begin{equation}
\frac{A_{ij}}{\Lambda} = Y^\Delta_{ij} \frac{g_{\Delta \phi}}{m_{\Delta}^2} \,, 
\label{ec:3.2b}
\end{equation}
with $m_\Delta$ the mass of the triplet.

Another possibility is the exchange of heavy right-handed neutrinos $\nu_R$ (the
seesaw mechanism). In this case, the relevant terms are
\begin{equation}
\mathcal{L}_{\nu_R} = - Y^\nu_{ij} \, \bar \ell_{Li} \, i \sigma_2 \,
\phi^* \, \nu_{Rj}
- \frac{1}{2} M_{Rij} \, \overline{\nu_{Ri}^c} \nu_{Rj} + \mathrm{H.c.} \,,
\label{ec:3.3}
\end{equation}
with $Y^\nu$ is the Dirac neutrino Yukawa coupling matrix
and $M_R$ the heavy right-handed neutrino mass matrix.
The integration of the heavy neutrino fields generates an effective
mass operator for the light neutrinos of the form
\begin{equation}
\frac{A_{ij}}{\Lambda} = - \frac{1}{2} (Y^\nu M_R^{-1} Y^{\nu T})_{ij} \,.
\label{ec:3.3b}
\end{equation}

After EWSB, the terms in Eq.~(\ref{ec:3.1}) yield the mass terms for the charged
leptons and left-handed neutrinos. Using matrix notation in flavor space,
the mass terms can be written as
\begin{equation}
\mathcal{L}_m = -\bar e_L M_e e_R  - \frac{1}{2} \bar \nu_L M_L \nu_L^c \,,
\label{ec:3.4}
\end{equation}
where $M_e = v \, Y^e$ and $M_L = 2 A \,v^2/\Lambda$, with $v=174$ GeV the
vacuum expectation value (VEV) of the SM Higgs boson.

Our parametrization of the new physics effects is obtained after general
considerations on weak basis independence. For convenience, let us define
\begin{equation}
\hat M_L \equiv \varepsilon \, A = \frac{\varepsilon \Lambda}{2 v^2} \, M_L 
\equiv \frac{M_L}{\mathcal{N}} \,,
\label{ec:3.4b}
\end{equation}
with $\varepsilon$ a dimensionless parameter to be
specified later. Under the change of weak basis
\begin{equation}
\ell_L  = V_L ~ \ell'_L  \,,\quad e_R = V_R^e ~ e'_R
\label{ec:3.5}
\end{equation}
we have
\begin{equation}
Y^e \to V_L^\dagger \, Y^e \, V_R^e \,,\quad
\hat M_L \to V_L^\dagger \hat M_L V_L^* \,.
\label{ec:3.6}
\end{equation}
Let us assume that some perturbations arising from radiative corrections are
added to the ``tree-level'' matrices $Y^e$ and $\hat M_L$,
\begin{equation}
Y^e \to Y^e + \delta Y^e \,,\quad \hat M_L \to \hat M_L + \delta \hat M_L \,.
\label{ec:3.7}
\end{equation}
The matrices $\delta Y^e$, $\delta \hat M_L$ are functions of $Y^e$, $\hat M_L$
and other SM and new physics parameters.
Under the change of basis defined in Eqs.~(\ref{ec:3.5}), the perturbations
must transform as
\begin{equation}
\delta Y^e \to V_L^\dagger \, \delta Y^e \, V_R^e \,,\quad
\delta \hat M_L \to V_L^\dagger \, \delta \hat M_L \, V_L^* \,,
\label{ec:3.8}
\end{equation}
since the physical observables must be independent of the choice of weak
basis.\footnote{These transformation properties for $\delta Y^e$ and
$\delta \hat
M_L$ do not assume that the Lagrangian is invariant under the change of basis
in Eqs.~(\ref{ec:3.5}) alone. Within the SM, the Lagrangian is invariant under
these transformations, but this does not happen in some of its extensions, for
instance in the MSSM. Besides, at very high energies, some flavor symmetry
might single out a special weak basis. Below that scale, and in particular at
low energies, this symmetry is broken.}
These transformation laws imply that the perturbations have the form
\cite{anarch}
\begin{eqnarray}
\delta Y^e & = & \lambda_e \, Y^e
+ \zeta_e \, Y^e Y^{e \dagger} Y^e
+ \eta_e \, \hat M_L \hat M_L^\dagger Y^e
+ \cdots \,, \nonumber \\
\delta \hat M_L & = & \lambda_L \, \hat M_L 
+ \zeta_L \, \hat M_L \hat M_L^\dagger \hat M_L \nonumber \\
& & + \eta_L \left( Y^e Y^{e \dagger} \hat M_L 
  + \hat M_L Y^{e*} Y^{eT} \right) + \cdots \,,
\label{ec:3.9}
\end{eqnarray}
where the $\lambda_i$, $\zeta_i$ and $\eta_i$ coefficients ($i=e,L$) are
functions of $Y^e$, $\hat M_L$ and of the coupling constants, which are
invariant under the transformations of Eqs.~(\ref{ec:3.5}) and in general
complex. The higher order terms in this expansion are expected to be
smaller.
The effect of the $\lambda_i$ terms in Eqs.~(\ref{ec:3.9}) is to rescale
the masses by common factors $(1+\lambda_e)$ for charged leptons and
$(1+\lambda_L)$ for neutrinos, without affecting neither the mass hierarchy
nor the
mixing. The $\zeta_i$ terms also rescale the masses, but with a different
factor for each lepton:
$m_{e_j} \to m_{e_j} \, (1+\zeta_e \,m_{e_j}^2/v^2)$ for charged leptons and
$m_j \to m_j \, (1+\zeta_L \,m_j^2/\mathcal{N}^2)$ for neutrinos. Hence, the
$\zeta_i$ terms modify the mass hierarchies. The $\eta_i$ terms are the
lowest-order ones which modify the leptonic mixing.

For some SM extensions, in principle there may exist a matrix $X$ (not
necessarily
square) of couplings between the leptons and other particles,
transforming as $Y^e$ or $\hat M_L$ either on the left or on the right side.
For instance, if $X$ transforms under the change of basis in
Eqs.~(\ref{ec:3.5}) as
\begin{equation}
X \to V_L^\dagger \, X \, V_R^X \,,
\label{ec:3.10}
\end{equation}
this matrix would contribute to $\delta Y^e$ and $\delta \hat M_L$ with terms
$X X^\dagger Y^e$ and $(X X^\dagger \hat M_L + \hat M_L X^* X^T)$, respectively.
Such possibility will not be considered in the following, and in this respect
our analysis is not be the most general one.\footnote{Due to the ignorance
regarding the structure of this matrix $X$, the
discussion of the effects of the terms $X X^\dagger Y^e$,
$(X X^\dagger \hat M_L + \hat M_L X^* X^T)$ is not feasible.}
We thus assume $Y^e$ and $\hat M_L$ as
the only sources of flavor violation in the lepton sector.
This case can naturally arise if some symmetry relates the couplings in
Eq.~(\ref{ec:3.1}) with the ones between the leptons and the new particles.

\begin{figure}[hbt]
\begin{center}
\begin{tabular}{ccc}
\mbox{\epsfig{file=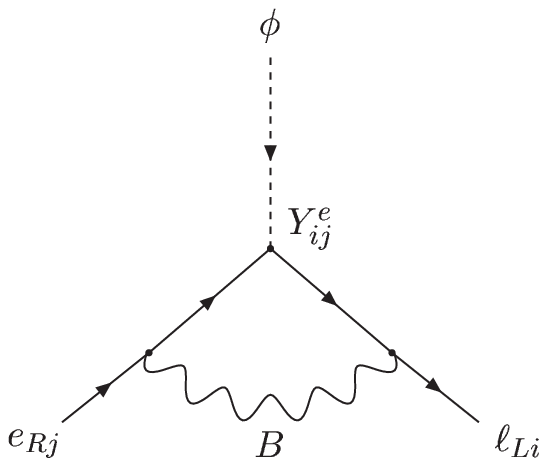,width=3.5cm,clip=}} & ~~~ &
\mbox{\epsfig{file=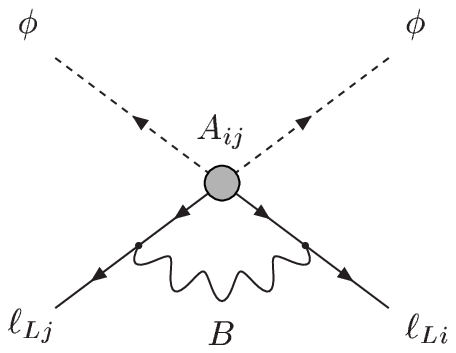,width=3cm,clip=}} \\[0.2cm]
(a) & ~~~ & (b) \\[0.5cm]
\mbox{\epsfig{file=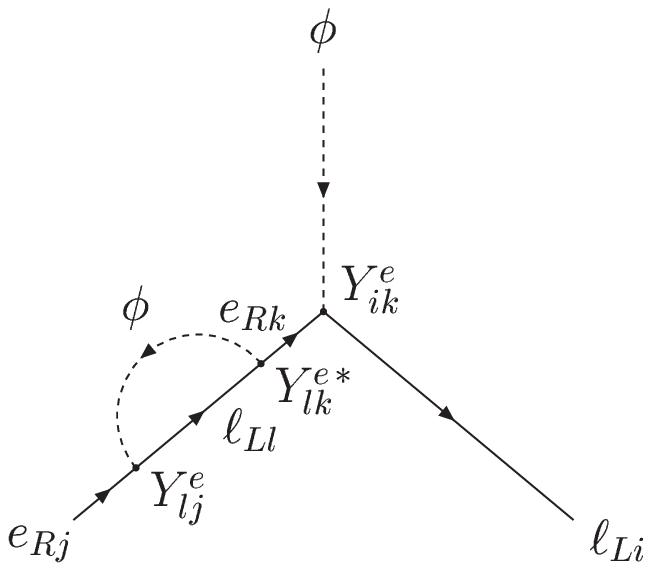,width=3.5cm,clip=}} & ~~~ &
\mbox{\epsfig{file=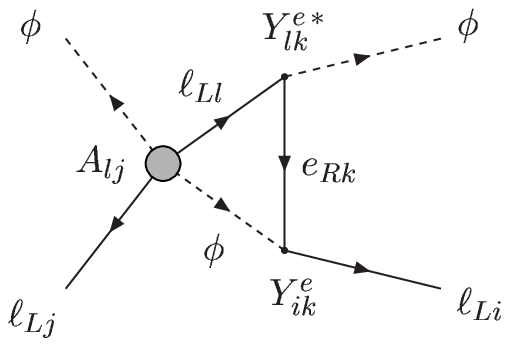,width=3.5cm,clip=}}
 \\[0.2cm]
(c) & ~~~ & (d) \\[0.5cm]
\mbox{\epsfig{file=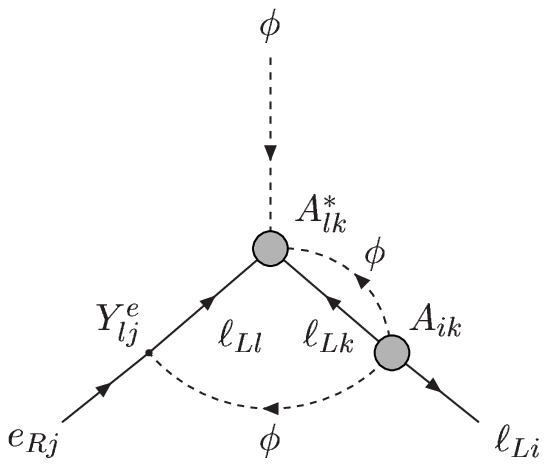,width=3.5cm,clip=}} & ~~~ &
\mbox{\epsfig{file=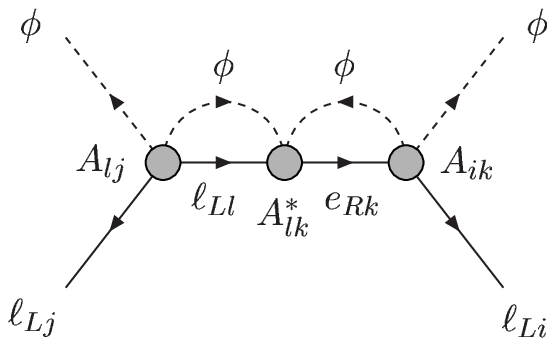,width=3.5cm,clip=}} \\[0.2cm]
(e) & ~~~ & (f)
\end{tabular}
\end{center}
\caption{Examples of SM diagrams giving corrections to the
$\ell_L \phi \, e_R$ and $\ell_L \ell_L^c \phi^* \phi^*$ vertices.
In each vertex the flavor dependence is explicitly shown.
\label{fig:diags}}
\end{figure}

It is worthwhile showing some examples of Feynman diagrams which
contribute to the different terms in Eqs.~(\ref{ec:3.9}) within the SM,
including also the effective neutrino mass operator in Eq.~(\ref{ec:3.1}) as
part of the SM vertices. The $\lambda_i$ terms result
from diagrams with minimal flavor structure, as for example diagrams (a) and
(b) in Fig.~\ref{fig:diags}, with the exchange of a $B$ boson with
flavor-universal couplings. The remaining
terms in Eqs.~(\ref{ec:3.9}) require the exchange of one or more $\phi$
doublets. In particular, the $\zeta_e$ and $\eta_L$ terms arise from diagrams
like (c) and (d) in Fig.~\ref{fig:diags}, respectively. 
At the one loop level the terms with $\eta_e$
and $\zeta_L$ are absent, and to generate them it is
necessary to consider two-loop corrections, for instance diagrams (e) and (f),
respectively.

In new physics scenarios there are additional interactions that may or may not
be suppressed by a large scale $\Lambda'$. These interactions mediate Feynman
diagrams giving further corrections to the
$\ell_L \phi \, e_R$ and $\ell_L \ell_L^c \phi^* \phi^*$ vertices.
Several examples of new physics contributions to these operators can be
found in Ref.~\cite{smirnov}.
We remark that, if the particle content and parameters of the new physics model
are specified, the corrections to the charged lepton and neutrino mass matrices
can be completely determined. However, in the absence of any experimental
indication favoring any of the theories
beyond the SM, the $\lambda_i$, $\zeta_i$, $\eta_i$ coefficients cannot be
predicted. Then, it is sensible to perform a statistical analysis in order to
determine, under certain assumptions, how much the tree-level predictions can
change due to radiative corrections from new physics, parametrized according to
Eq.~(\ref{ec:3.9}), leaving $\lambda_i$, $\zeta_i$, $\eta_i$ as unknown
parameters.
The size of these coefficients is expected to be
similar, because all the terms in Eqs.~(\ref{ec:3.9}) can be generated at
the one loop level (although the terms with $\eta_e$ and $\zeta_L$ appear at
next-to-leading order within the SM, they may arise at leading order in other
models, e.g. with a scalar triplet). On the other hand, the
higher order terms omitted in Eqs.~(\ref{ec:3.9}), involving products of 5
matrices or more, are expected to be suppressed
by a factor $\sim 10$ with respect to the leading ones.

One crucial issue for our analysis is the value of the parameter $\varepsilon$
in Eq.~(\ref{ec:3.4b}), which accounts for the normalization of $\hat M_L$.
The size of this parameter reflects the suppression of the new interactions,
and determines the relative importance of $\hat M_L$ with
respect to $Y^e$ in the expressions used for the perturbations. We consider two
limiting scenarios:

(i) In the first scenario we take
$\varepsilon = 1$, in which case $\hat M_L = A$ with matrix elements of order
unity. This corresponds to a situation where there are new interactions which
are not suppressed by a large scale $\Lambda'$.
In this scenario, the terms
$\zeta_L \, \hat M_L \hat M_L^\dagger \hat M_L$ and
$\eta_e \, \hat M_L \hat M_L^\dagger Y^e$ in Eqs.~(\ref{ec:3.9})
are not negligible, and have an
important influence on the neutrino mass hierarchy and mixing,
respectively.

(ii) In the second scenario we assume $\varepsilon \ll 1$, so that
these two terms (which have two or more powers of $\hat M_L$) can be omitted in
Eqs.~(\ref{ec:3.9}), resulting in
\begin{eqnarray}
\delta Y^e |_\mathrm{S2} & = & \lambda_e \, Y^e
+ \zeta_e \, Y^e Y^{e \dagger} Y^e
+ \cdots \,, \nonumber \\
\delta \hat M_L |_\mathrm{S2} & = & \lambda_L \, \hat M_L 
+ \eta_L \left( Y^e Y^{e \dagger} \hat M_L 
+ \hat M_L Y^{e*} Y^{eT} \right) \nonumber \\
& & + \cdots \,.
\label{ec:3.11}
\end{eqnarray}
This scenario corresponds to a new physics model in which the new interactions
are suppressed by a large scale $\Lambda' \sim \Lambda$, for instance in models
based in the seesaw mechanism.
In the limit $\varepsilon \ll 1$, the normalization of $\hat M_L$ is irrelevant.

From Eqs.~(\ref{ec:3.11}) we see that in scenario 2
the expressions for the perturbations are formally similar to
the one loop RG equations within the SM. Therefore, in this scenario the results
obtained within our framework are expected to be similar to the results of RG
evolution, bearing in mind that in the case of the RG
equations the coefficients $\lambda_i$, $\zeta_e$ and $\eta_L$ are fixed, while
in our case they are unknown in principle.

If the neutrinos are not Majorana but Dirac particles, the Yukawa term of the
Lagrangian originating their masses reads
\begin{equation}
\mathcal{L}_D = - Y^\nu_{ij} \, \bar \ell_{Li} \, i \sigma_2 \phi^* \, \nu_{Rj}
+ \mathrm{H.c.}
\label{ec:3.12}
\end{equation}
For this term and the charged lepton one, the change of basis analogous to
Eqs.~(\ref{ec:3.5}) reads
\begin{equation}
\ell_L  = V_L ~ \ell'_L  \,,\quad e_R = V_R^e ~ e'_R \,,
\quad \nu_R = V_R^\nu ~ \nu'_R \,.
\label{ec:3.13}
\end{equation}
Under this transformation, the Yukawa matrices transform as
\begin{equation}
Y^e \to V_L^\dagger \, Y^e \, V_R^e \,,\quad
Y^\nu \to V_L^\dagger \, Y^\nu \, V_R^\nu \,.
\label{ec:3.14}
\end{equation}
This allows to obtain the expressions for the perturbations for Dirac
neutrinos,
\begin{eqnarray}
\delta Y^e & = & \lambda_e \, Y^e
+ \zeta_e \, Y^e Y^{e \dagger} Y^e
+ \eta_e \, Y^\nu \hat Y^{\nu \dagger} Y^e
+ \cdots \,, \nonumber \\
\delta Y^\nu & = & \lambda_\nu \, Y^\nu
+ \zeta_\nu \, Y^\nu Y^{\nu \dagger} Y^\nu 
+ \eta_\nu \, Y^e Y^{e \dagger} Y^\nu \nonumber \\
& & + \cdots \,.
\label{ec:3.15}
\end{eqnarray}
In the following, we will generally refer to the case where neutrinos are
Majorana particles, and quote the results for Dirac neutrinos when
relevant.

\section{Numerical results}
\label{sec:4}

Using the parametrization of new physics contributions given in
Eqs.~(\ref{ec:3.9}), (\ref{ec:3.11}), (\ref{ec:3.15}), we study the changes in
the pattern of neutrino masses and mixings due to these corrections.
For this purpose, we take
the unknown coefficients $\lambda_i$, $\zeta_i$, $\eta_i$ as random complex
parameters, generated with a Gaussian
distribution centered at zero and, for simplicity, we assume that the standard
deviations coincide:
\begin{equation}
\langle |\lambda_i|^2 \rangle^\frac{1}{2} =
\langle |\zeta_i|^2 \rangle^\frac{1}{2} = 
\langle |\eta_i|^2 \rangle^\frac{1}{2} \equiv \kappa \,.
\label{ec:4.1}
\end{equation}
Contrarily to what could be expected,
this is not a serious bias in the analysis, because the moduli of the random
parameters
are not constrained to be all equal, and only the standard deviations of the
distributions are assumed to be the same. The phases of $\lambda_i$, $\zeta_i$
and $\eta_i$ are generated uniformly
between $0$ and $2\pi$. In our study the following procedure is
applied: we fix a value of $\kappa$ and generate a large set of matrices
using Eqs.~(\ref{ec:3.9}), (\ref{ec:3.11}) or
(\ref{ec:3.15}), as appropriate, with
random coefficients $\lambda_i$, $\zeta_i$, $\eta_i$. These
matrices are diagonalized in order to obtain the masses and the
neutrino mixing matrix. We then select some observable, and examine its
distribution over the set of matrices. For each value of $\kappa$, the
$1 \sigma$ limits on
this observable are defined as the boundaries of the 68.3\% confidence level
central interval, evaluated from the sample of random matrices. These
$1 \sigma$ limits reflect the ``average'' behavior of the observable under
consideration, when arbitrary perturbations are added to the original matrices.
It must be emphasized that the maximum and minimum values can be very different
from the average values, and some situations are found where in average the
observable does not change appreciably under perturbations, but for
fine-tuned values of the random parameters it does.

In the numerical analysis we take initial ``tree-level'' matrices $Y^e$ and
$\hat M_L$ ($Y^\nu$ for Dirac neutrinos)
reproducing the current experimental data summarized in Section \ref{sec:2}.
The charged lepton masses are taken at the scale $M_Z$ \cite{koide}. We assume
$\sin \theta = 0.15$ 
and fix the Dirac and Majorana phases to be $\delta = \pi/2$,
$\alpha = \pi/3$ and $\beta = \pi/5$. 
We analyze separately the two possibilities of a HI or a QD spectrum.
For the
case of an inverted hierarchy the results turn out to be very similar to those
found for a normal hierarchy, and we do not present them.
We summarize our input values in Table~\ref{tab:1}.
In scenario 1, we take $\hat M_L$ with the mass of the heaviest
neutrino normalized to unity. The normalization of $\hat M_L$ is
irrelevant in scenario 2, as shown in the previous section.

\begin{table}
\begin{tabular}{cc}
\hline
\hline
Parameter & Value \\
\hline
$m_e$ & $0.487$ MeV \\
$m_\mu$ & 0.103 GeV \\
$m_\tau$ & 1.747 GeV \\
$m_1$ & $\left\{
\mbox{\begin{tabular}{l} $10^{-5}$ eV (HI) \\ 1 eV (QD) \end{tabular}}
 \right.$ \\
$\Delta m_\odot^2$ & $7.13 \times 10^{-5}$ eV \\
$\Delta m_\mathrm{atm}^2$ & $2.6 \times 10^{-3}$ eV \\
$\tan \theta_\odot$ & 0.62 \\
$\tan \theta_\mathrm{atm}$ & 1 \\
$\sin \theta$ & 0.15 \\
$\delta$ & $\pi/2$ \\
$\alpha$ & $\pi/3$ \\
$\beta$ & $\pi/5$ \\
\hline
\hline
\end{tabular}
\caption{Input parameters used for the unperturbed mass matrices.
\label{tab:1}}
\end{table}

\subsection{Stability of mass and mixing parameters}
\label{sec:4.1}

Let us discuss how the parameters $r \equiv \Delta m^2_\odot/\Delta
m^2_\mathrm{atm}$, $\tan \theta_\odot$, $\tan \theta_\mathrm{atm}$ and
$\sin \theta$ change when corrections are added to the mass
matrices. Our aim is to investigate whether these quantities are stable or not,
and to what extent they get modified by the perturbations.
We first present the results for
scenario 1, and later discuss the differences with scenario 2 and the results
for Dirac neutrinos.

\begin{figure}[htb]
\begin{center}
\epsfig{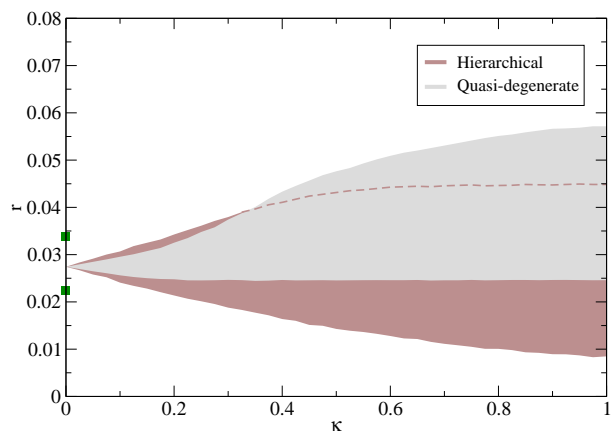}
\end{center}
\caption{Effect of the perturbations on the ratio $r$ in scenario 1. For
illustration, the present $1\sigma$ limits are displayed on the vertical axis.
\label{fig:R}}
\end{figure}

Regarding the ratio of mass squared differences $r$, we observe in
Fig.~\ref{fig:R} that the
corrections to the matrices have a large impact on this quantity, both in the
cases of a HI or QD spectrum. This plot (and the remaining ones
in this section) must be interpreted with caution: it does not
provide any limit on the size of the corrections, on the basis of the
experimental measurement of $r$, because the initial ``tree-level'' value we
use for $r$ needs not be equal to the observed value. Instead, the meaning of
the plot
is that $r$ is not stable under perturbations, and from an initial value chosen
to be $r=0.027$ one can obtain values between 0.021 and 0.034, for a
HI spectrum and $\kappa = 0.2$.

\begin{figure}[htb]
\begin{center}
\epsfig{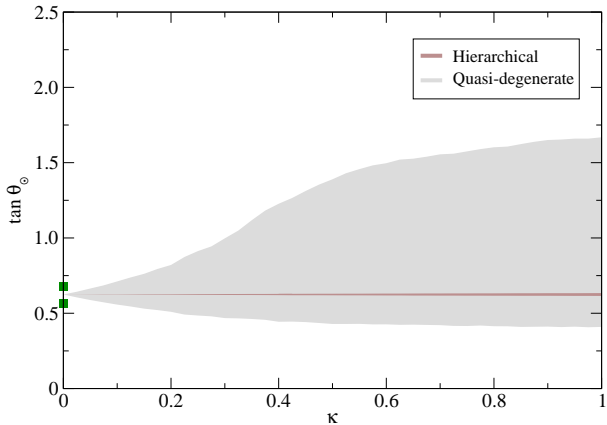}
\end{center}
\caption{Effect of the perturbations on $\tan \theta_\odot$ in scenario 1. For
illustration, the present $1\sigma$ limits are displayed on the vertical axis.
\label{fig:tan12}}
\end{figure}

The effect of the perturbations on $\tan \theta_\odot$ is quite different for
a HI or QD spectrum. In the former case, $\tan
\theta_\odot$ is very stable even for relatively large perturbations, as can be
noticed in
Fig.~\ref{fig:tan12}. On the contrary, for quasi-degenerate neutrinos, the value
of $\tan \theta_\odot$ can change significantly with new physics corrections.
This fact suggests that, if neutrinos are quasi-degenerate, the underlying
tree-level pattern of lepton mass matrices could
correspond to bi-maximal mixing, being the observed value
$\tan \theta_\odot \simeq 0.6$
the result of radiative corrections. We will analyze in detail this possibility
at the end of this section.

\begin{figure}[htb]
\begin{center}
\epsfig{file=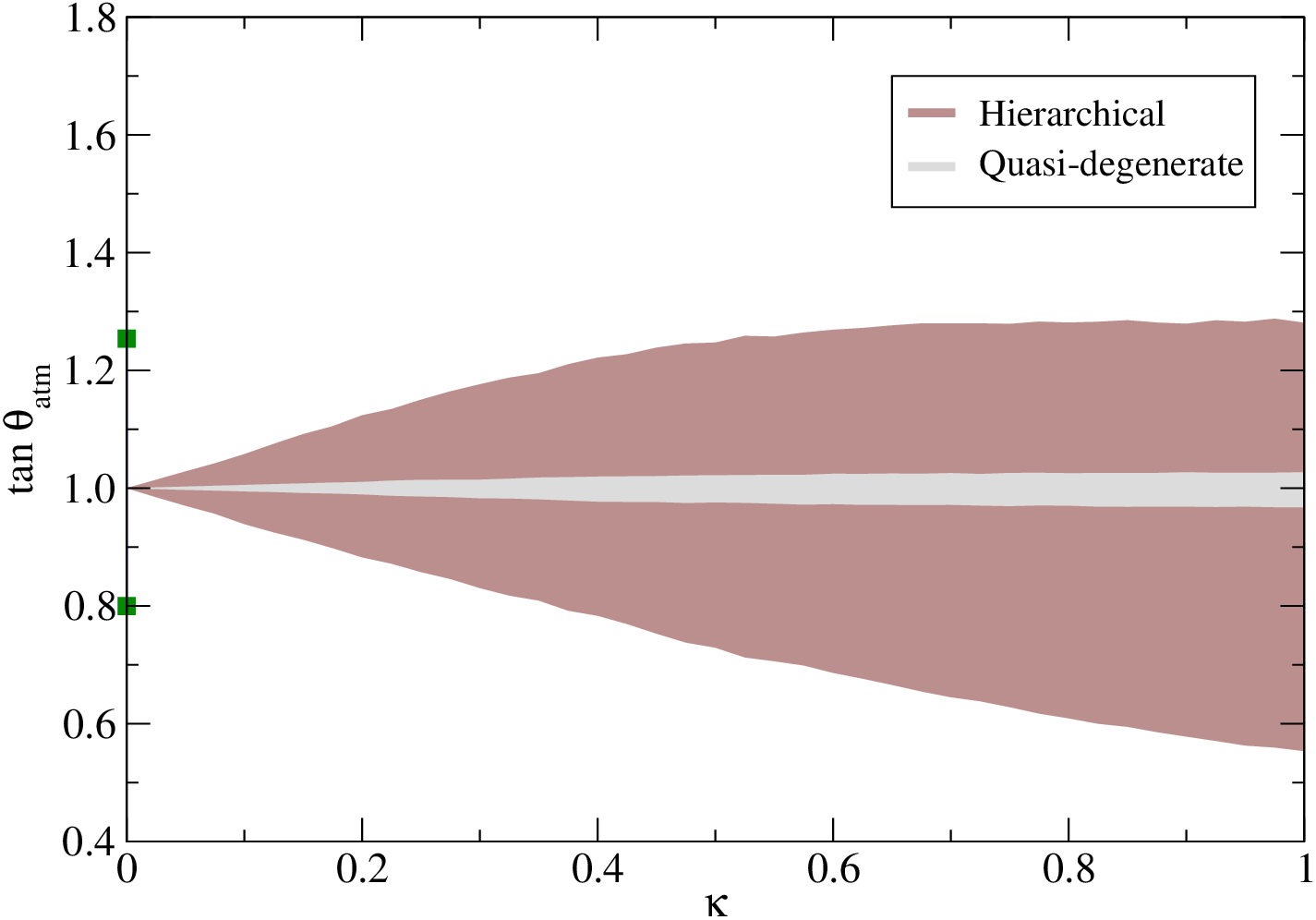,width=8cm,clip=}
\end{center}
\caption{Effect of the perturbations on $\tan \theta_\mathrm{atm}$ in scenario
1. For illustration, the present $1\sigma$ limits are displayed on the
vertical axis.
\label{fig:tan23}}
\end{figure}

The behavior of $\tan \theta_\mathrm{atm}$ is the opposite to the one observed
for $\tan \theta_\odot$, as it can be perceived from
Fig.~\ref{fig:tan23}: for a HI spectrum this parameter is modified
by perturbations on the mass matrices, while for a
QD spectrum it is fairly stable. This shows that, for the
case of quasi-degenerate neutrinos, the experimental observation of
$\tan \theta_\mathrm{atm} \simeq 1$ must correspond to $\tan
\theta_\mathrm{atm} \simeq 1$ in the mass
matrices, because this prediction is not altered by the corrections. On the
other hand, for a HI spectrum the observation of
$\tan \theta_\mathrm{atm} \simeq 1$ could be either a coincidence, or result
from a specific symmetry leading naturally to this value and making higher
order corrections very small.

\begin{figure}[htb]
\begin{center}
\epsfig{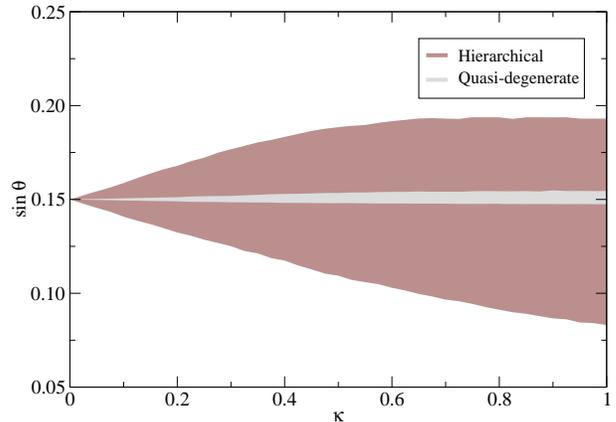}
\end{center}
\caption{Effect of the perturbations on $\sin \theta$ in scenario 1.
\label{fig:V13}}
\end{figure}

\begin{figure}[htb]
\begin{center}
\epsfig{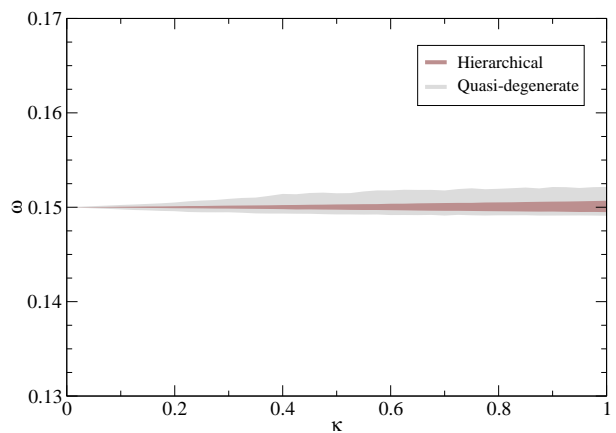}
\end{center}
\caption{Effect of the perturbations on the ratio $\omega=\sin \theta/\tan
\theta_\mathrm{atm}$ in scenario 1.
\label{fig:omega}}
\end{figure}

The analysis of $\sin \theta$ (which equals $|U_{e3}|$ for a normal hierarchy)
shows that it does not
change under perturbations for a QD spectrum, but it is considerably modified
when the neutrino masses are hierarchical (see Fig.~\ref{fig:V13}). However,
one striking feature of our analysis is that the ratio
\begin{equation}
\omega = \frac{\sin \theta}{\tan \theta_\mathrm{atm}}
\label{ec:4.2}
\end{equation}
remains practically constant even when large perturbations are added to $Y^e$
and $\hat M_L$: for $\kappa = 1$ $\sin \theta$ and $\tan \theta_\mathrm{atm}$
change by
more than $\pm 40$\%, while their ratio changes less than 1\%, as can be
noticed in Fig.~\ref{fig:omega}.

This feature may have important consequences for model building. When
$\sin \theta$ is experimentally measured, the ratio $\omega$ will be an
excellent tool to investigate the structure of the lepton mass matrices, because
it is very insensitive to radiative corrections (as long as these corrections
do not have any additional source of flavor violation, which is the framework
we consider).
This ratio will then allow to test experimentally, with a ``clean'' observable,
the textures for lepton mass
matrices implied by flavor symmetries proposed in the literature.

The remaining parameters to be investigated are the $CP$ violating phases, for
which the results depend on the initial values used (see
Table~\ref{tab:1}). For a
HI spectrum, the Dirac phase $\delta$ remains virtually constant at its initial
value, while the Majorana phases $\alpha$ and $\beta$ vary over a
wide range, $0.6 \leq \alpha \leq 2$, $0.4 \leq \beta \leq 2.2$ for
$\kappa = 1$. For a QD spectrum, the Majorana phases
remain within $\pm 10$\% of their initial value for $\kappa = 1$, while
the Dirac phase varies
in the interval $1 \leq \delta \leq 1.8$.

In scenario 2, the behavior is very different for a HI spectrum. In this case,
we find
that $r$, $\tan \theta_\odot$, $\tan \theta_\mathrm{atm}$, $\sin \theta$ and
the three $CP$
violating phases remain constant when the perturbations in Eqs.~(\ref{ec:3.11})
are added to the matrices. In the SM these
quantities exhibit a similar behavior under RG
evolution \cite{RGevol},
as its expressions are formally identical to ours. On the other hand,
in the
case of a QD spectrum, the differences between scenarios 1 and 2 are not
significant, and the discussion above applies also to scenario 2. This contrast
can be understood in view of the analysis of the dependence on the
neutrino masses presented in Section \ref{sec:4.2} below.

For Dirac neutrinos, the results are found to be rather similar to
the ones obtained in scenario 2. For a HI spectrum the parameters under
consideration do not change when the perturbations in Eqs.~(\ref{ec:3.15}) are
included. For a QD spectrum, however, they are modified, following practically
the same behavior that can be observed in the plots for scenario 1 shown in
this section.

Finally, we note that the results on $r$, $\tan \theta_\odot$ and $\tan
\theta_\mathrm{atm}$ are
almost independent on the value of $\sin \theta$ used. Other quantities
obviously depend on the particular value of $\sin \theta$, as for example
$\mathcal{J}$, which is proportional to $\sin \theta$.
The choice of the $CP$-violating phases is only relevant for
the ``tree-level'' values of quantities that depend on them, like $\mathcal{J}$
and $m_{ee}$. For a QD spectrum, the deviations of $r$, $\tan \theta_\odot$,
$\tan \theta_\mathrm{atm}$, $\sin \theta$ and $\omega$ are very similar, and
for a HI spectrum the influence of phases on these quantities is completely
negligible. We have also checked that our results do not change when the
next terms in the expansion of Eqs.~(\ref{ec:3.9}) (with products of 5
matrices) are included, even in the unrealistic limit where these terms
have similar coefficients. We have found that the quantities that are stable
remain stable, and the quantities that change under perturbations exhibit an
analogous behavior with the inclusion of these terms.

\subsection{Dependence on the neutrino masses}
\label{sec:4.2}

\begin{figure}[htb]
\begin{center}
\epsfig{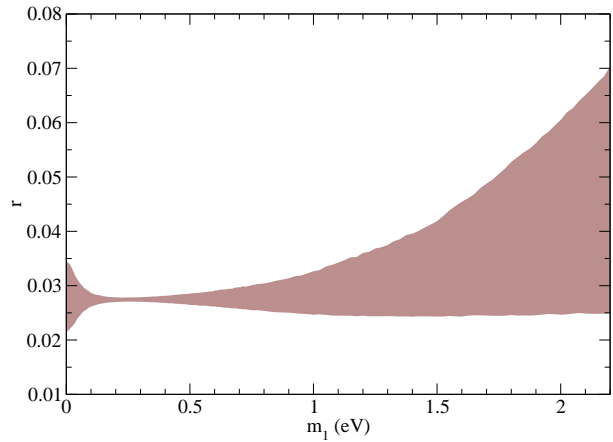}
\end{center}
\caption{Effect of the perturbations on $r$ for $\kappa = 0.2$, as a function
of the mass of the lightest neutrino, in scenario 1.
\label{fig:R-m}}
\end{figure}

\begin{figure}[htb]
\begin{center}
\epsfig{file=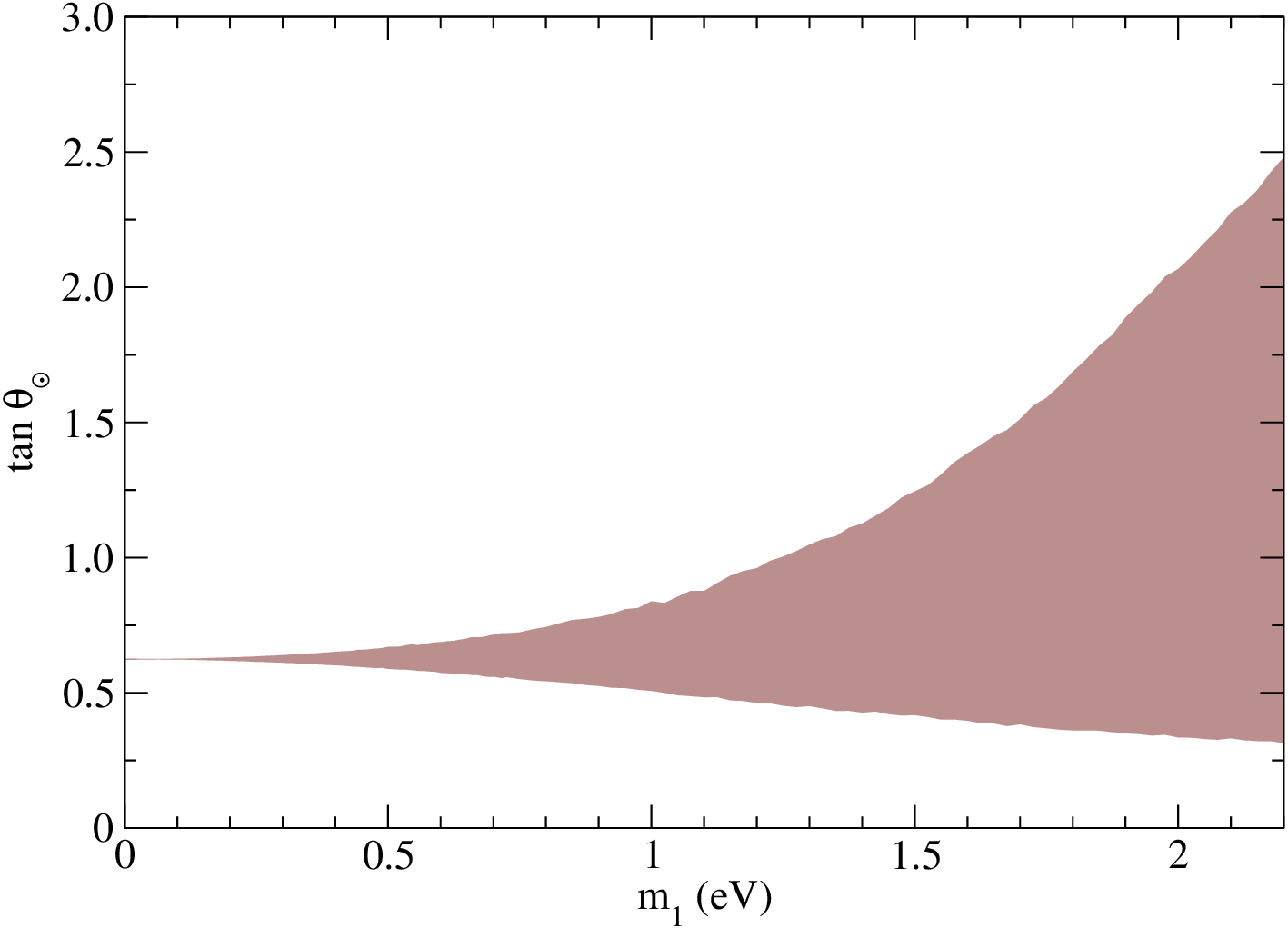,width=8cm,clip=}
\end{center}
\caption{Effect of the perturbations on $\tan \theta_\odot$ for $\kappa = 0.2$,
as a function of the mass of the lightest neutrino, in scenario 1.
\label{fig:tan12-m}}
\end{figure}

\begin{figure}[htb]
\begin{center}
\epsfig{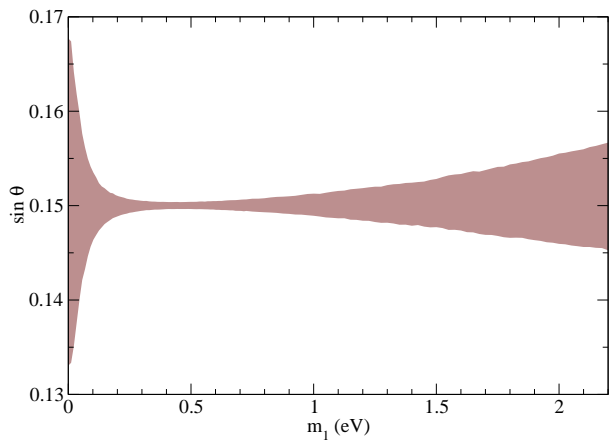}
\end{center}
\caption{Effect of the perturbations on $\sin \theta$ for $\kappa = 0.2$,
as a function of the mass of the lightest neutrino, in scenario 1.
\label{fig:V13-m}}
\end{figure}

We have verified that the influence of the perturbations on some parameters
depends
strongly on the type of neutrino spectrum, namely the deviations of $\tan
\theta_\odot$ are negligible for a HI spectrum while they are large if the
neutrinos are quasi-degenerate. It is then convenient to analyze the dependence
of the deviations on the mass of the lightest neutrino, which we take between
$10^{-5}$ eV and 2.2 eV (the direct bound from the Mainz and Troitsk
experiments). In Figs.~\ref{fig:R-m}--\ref{fig:V13-m} we plot the
effect of the corrections on $r$, $\tan \theta_\odot$, $\tan
\theta_\mathrm{atm}$ and
$\sin \theta$, respectively, for $\kappa = 0.2$.

\begin{figure}[htb]
\begin{center}
\epsfig{file=Figs/tan23-m.eps,width=8cm,clip=}
\end{center}
\caption{Effect of the perturbations on $\tan \theta_\mathrm{atm}$ for
$\kappa = 0.2$, as a function of the mass of the lightest neutrino, in scenario
1.
\label{fig:tan23-m}}
\end{figure}

It is apparent that these parameters are remarkably more stable in the region
around $m_1 = 0.3$ eV than for the rest of values of $m_1$. This can be
understood as follows: the only two terms that influence
the mixing are $\eta_e \, \hat M_L \hat M_L^\dagger Y^e$ and
$\eta_L \left( Y^e Y^{e \dagger} \hat M_L + \hat M_L Y^{e*} Y^{eT} \right)$. Of
these, the former is relevant only for a HI spectrum, whereas for a QD spectrum
it does not have any influence. On the contrary, the latter term is important
for a QD spectrum but its impact is negligible if the neutrino masses are
hierarchical. Hence, the deviations in $\tan \theta_\mathrm{atm}$ and $\sin
\theta$ appearing in the left part of Figs.~\ref{fig:tan23-m}
and \ref{fig:V13-m} are due to the $\eta_e$ term, and the deviations in
$R$, $\tan \theta_\odot$, $\tan \theta_\mathrm{atm}$ and $\sin \theta$ that can
be seen
in the right side of Figs.~\ref{fig:R-m}--\ref{fig:V13-m} are a consequence
of the $\eta_L$ term. The deviations in $r$ in the left hand side of
Fig.~\ref{fig:R-m} is an effect of the $\zeta_L$ term, which does not contribute
to the mixing. 

\begin{figure}[htb]
\begin{center}
\epsfig{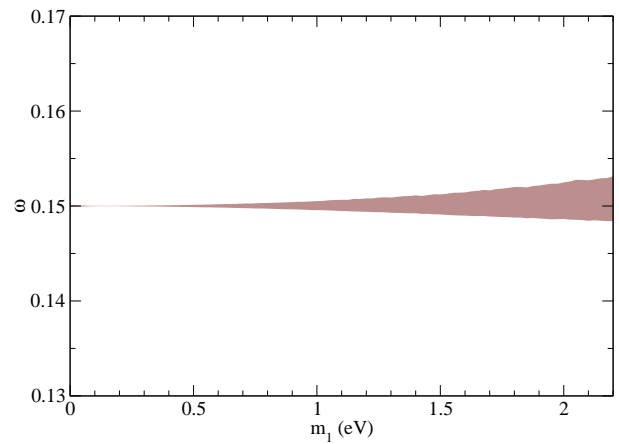}
\end{center}
\caption{Effect of the perturbations on $\omega$ for $\kappa = 0.2$,
as a function of the mass of the lightest neutrino, in scenario 1.
\label{fig:omega-m}}
\end{figure}

The region $m \simeq 0.3$ eV is of
special interest, since these neutrino masses will be probed in forthcoming
experiments like KATRIN, which is planned to start in 2007.
If the mass of the lightest neutrino happens to be in
this range, this will mean that the corrections to the tree-level mass matrices
will have a much smaller impact on the hierarchy of mass squared differences and
the mixing. The same is also true for the Dirac and Majorana phases. 
For completeness, in Fig.~\ref{fig:omega-m} we show the $1\sigma$ limits on
$\omega$ for the same range of $m_1$. We observe that this ratio remains
virtually constant in the whole interval.

For scenario 2 and for Dirac neutrinos, the dependence on the lightest neutrino
mass is much simpler. For a HI spectrum, all the quantities studied are stable,
and in these cases the effects of perturbations are as in
Figs.~~\ref{fig:R-m}--\ref{fig:omega-m} but without the deviations present in
the left part of some of these plots.

\subsection{Stability of flavor relations}
\label{sec:4.3}

\begin{figure}[htb]
\begin{center}
\epsfig{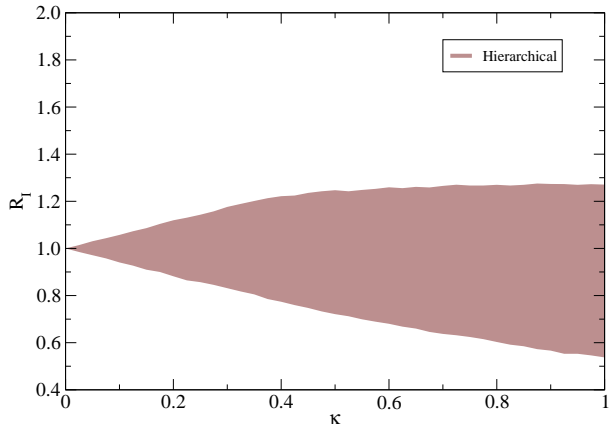}
\end{center}
\caption{Effect of the perturbations on the ratio $R_\mathrm{I}$, in scenario 1.
\label{fig:rel1}}
\end{figure}

\begin{figure}[htb]
\begin{center}
\epsfig{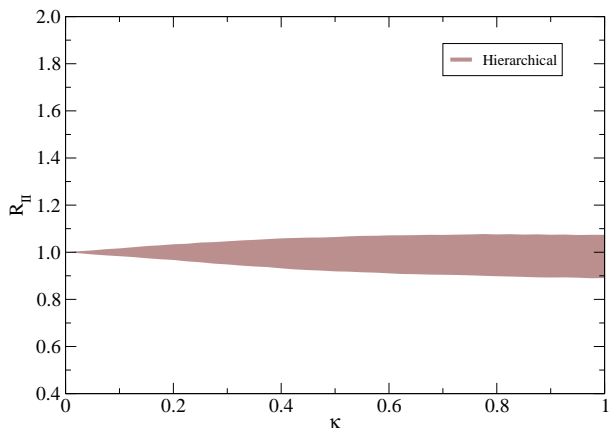}
\end{center}
\caption{Effect of the perturbations on the ratio $R_\mathrm{II}$, in
scenario 1.
\label{fig:rel2}}
\end{figure}

We are interested in finding flavor relations which are stable under
perturbations of the mass matrices. By ``stability'' we mean that, if these
relations hold for
the tree-level matrices, they still hold to a good approximation when
perturbations are added. With some exceptions, most
flavor relations found in the literature correspond to Majorana neutrinos
with a HI spectrum. In scenario 2, the parameters
$r$, $\tan \theta_\odot$, $\tan \theta_\mathrm{atm}$ and $\sin \delta$, as
well as the $CP$
violating phases, remain constant for a HI spectrum.
Therefore, any flavor relation among these parameters is stable. In
scenario 1, most of the flavor relations studied are affected by the random
perturbations. For illustration, we show the effect of perturbations on some
relations. For texture A1 in Ref.~\cite{barbieri}, there are two predictions,
\begin{eqnarray}
\sin \theta & = & \frac{1}{2} \tan \theta_\mathrm{atm} \sin 2 \theta_\odot
\sqrt r \,, \nonumber \\
\frac{|m_{ee}|}{m_\mathrm{atm}} & = & \sin^2 \theta_\odot \sqrt r \,,
\label{ec:rels1}
\end{eqnarray}
where $m_{ee}$ is the effective mass for
neutrinoless double beta decay processes, and $m_\mathrm{atm} \equiv
\sqrt{\Delta m_\mathrm{atm}^2}$. In order to test the stability of these
relations under the radiative corrections considered here, we
define the ratios
\begin{eqnarray}
R_\mathrm{I} & = & \frac{1}{2} \frac{\tan \theta_\mathrm{atm}}{\sin \theta}
\sin 2 \theta_\odot \sqrt r \,,
\nonumber \\
R_\mathrm{II} & = & \frac{\sin^2 \theta_\odot \sqrt
r}{|m_{ee}|/m_\mathrm{atm}} \,,
\end{eqnarray}
which equal unity for the tree-level matrices.
These ratios are plotted in Figs.~\ref{fig:rel1} and \ref{fig:rel2},
respectively, where we have used $\sin \theta = 0.074$, $\delta =
\alpha = 0$, $\beta = 0.85$, so that the initial matrices fulfill these
relations. The rest of parameters is taken from Table~\ref{tab:1}.

\begin{figure}[htb]
\begin{center}
\epsfig{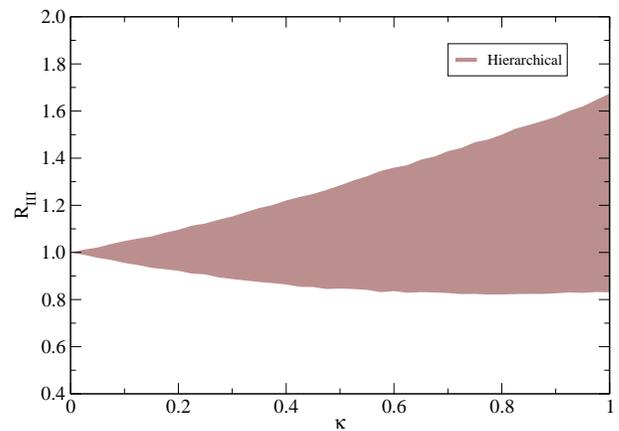}
\end{center}
\caption{Effect of the perturbations on the ratio $R_\mathrm{III}$, in
scenario 1.
\label{fig:rel3}}
\end{figure}

From Fig.~\ref{fig:rel1} we see that the first relation is modified when
corrections
are added to the mass matrices. At any rate, the deviations on this relation
are much smaller than the changes in $r$, $\tan \theta_\mathrm{atm}$ and
$\sin \theta$
(see Figs.~\ref{fig:R} -- \ref{fig:V13}). We also notice from
Fig.~\ref{fig:rel2} that the accuracy of
the second relation is hardly affected by perturbations on the mass matrices.
This feature makes its experimental test cleaner, and less dependent on unknown
corrections to the tree-level textures. For texture B1 of Ref.~\cite{barbieri}
(see also Ref.~\cite{tanimoto}) we have
\begin{equation}
\sin \theta = \frac{1}{2} \tan \theta_\mathrm{atm} \tan 2 \theta_\odot 
\sqrt{r \cos 2 \theta_\odot} \,.
\end{equation}
which differs from the first of Eqs.~(\ref{ec:rels1}) by factors depending
on $\theta_\odot$. Since $\theta_\odot$ is stable for a HI spectrum,
the effect of corrections on this relation is similar, and the plot obtained
is identical to Fig.~\ref{fig:rel1}. Another interesting relation is
\cite{ross}
\begin{equation}
\sin \theta = \sqrt{\frac{m_e}{m_\mu}} \,.
\end{equation}
The left-hand side of this equation varies with the perturbations, but the
right-hand side does not. We define
\begin{equation}
R_\mathrm{III}  = \frac{1}{\sin \theta} \sqrt{\frac{m_e}{m_\mu}}
\end{equation}
and set $\sin \theta = 0.0688$ (with the rest of parameters as in
Table~\ref{tab:1}) in order to test the stability of this relation. The result
can be seen in Fig.~\ref{fig:rel3}.

The conclusion  one may draw from the study of these examples is the
following: if the new physics interactions are suppressed (this situation
corresponds to scenario 2), the flavor relations are stable and, if they hold
at tree-level, they also hold when the corrections in Eqs.~(\ref{ec:3.11}) are
included. On the other hand, if the new interactions are not suppressed and the
corrections have the full form of Eqs.~(\ref{ec:3.9}), with $\hat M_L$ of order
unity (this
possibility corresponds to scenario 1) some of these flavor relations are
modified by the perturbations.

\subsection{Special limits}
\label{sec:4.4}

\begin{figure}[htb]
\begin{center}
\epsfig{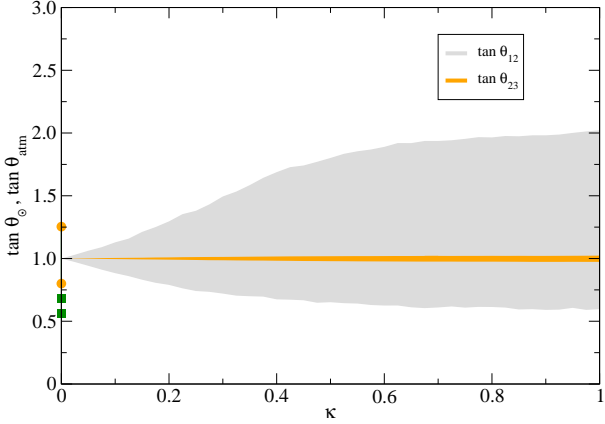}
\end{center}
\caption{Effect of the perturbations on $\tan \theta_\odot$ and $\tan
\theta_\mathrm{atm}$, for quasi-degenerate neutrinos and initial bi-maximal
mixing, in scenario 1.
\label{fig:bimax1}}
\end{figure}

\begin{figure}[htb]
\begin{center}
\epsfig{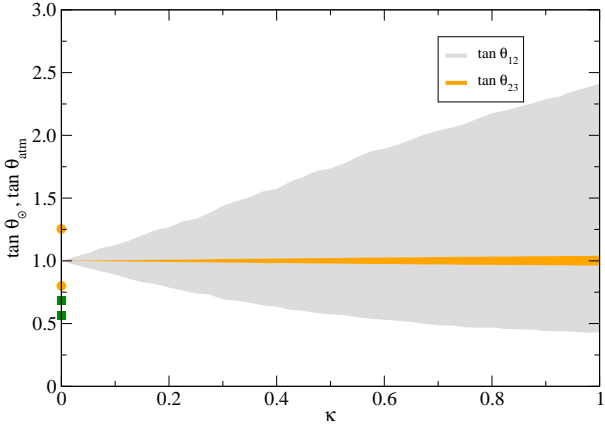}
\end{center}
\caption{Effect of the perturbations on $\tan \theta_\odot$ and $\tan
\theta_\mathrm{atm}$, for quasi-degenerate neutrinos and initial bi-maximal
mixing, in scenario 2.
\label{fig:bimax2}}
\end{figure}

We have previously remarked that, for quasi-degenerate neutrinos, the
perturbations in the mass matrices modify $\tan \theta_\odot$ but do not affect
$\tan \theta_\mathrm{atm}$
significantly. Then, one interesting question naturally arises: Is it possible
to have a QD spectrum with bi-maximal mixing at the tree level, so that the
smaller value of $\tan
\theta_\odot \simeq 0.6$ is due to effects of new physics? To test this
hypothesis, we set $\tan \theta_\odot = \tan \theta_\mathrm{atm} = 1$ in our
matrices, with the rest of parameters as in Table~\ref{tab:1}, and analyze
how $\tan \theta_\odot$ and $\tan \theta_\mathrm{atm}$ change when perturbations
are added.
For scenario 1, the results are displayed in Fig.~\ref{fig:bimax1}. For scenario
2, the results are shown in Fig.~\ref{fig:bimax2}.

From these figures we conclude that in both scenarios it is possible that,
from an initial bi-maximal pattern, large corrections to the mass matrices
modify significantly $\tan \theta_\odot$,
bringing it to its experimental value, while keeping $\tan \theta_\mathrm{atm}
\simeq 1$. This result is only slightly dependent on the value of $\sin \theta$
used, and is valid for $m_1 \gtrsim 0.6$ eV. For Dirac neutrinos the same effect
is found, and the results are very similar to the ones in scenario 2.

\begin{figure}[htb]
\begin{center}
\epsfig{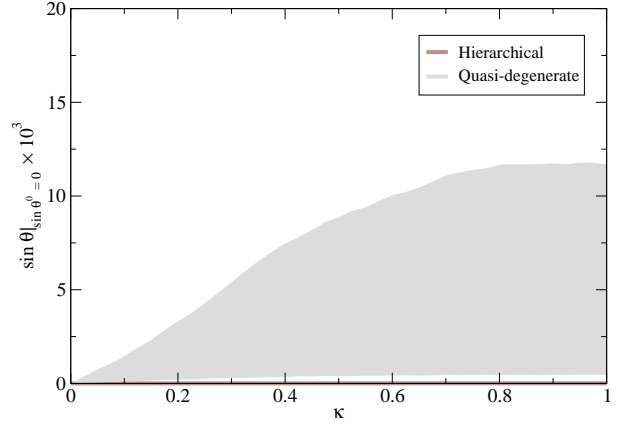}
\end{center}
\caption{Effect of the perturbations on $\sin \theta$,
for an initial vanishing value, in scenario 1.
\label{fig:V13-V0-1}}
\end{figure}

\begin{figure}[htb]
\begin{center}
\epsfig{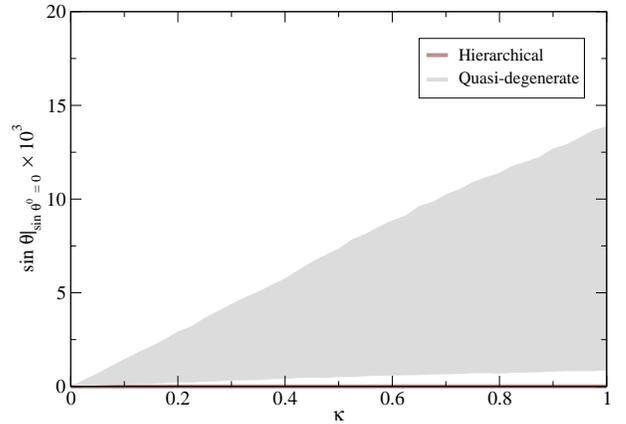}
\end{center}
\caption{Effect of the perturbations on $\sin \theta$,
for an initial vanishing value, in scenario 2.
\label{fig:V13-V0-2}}
\end{figure}

Other interesting situation corresponds to $\sin \theta = 0$ at the tree level.
In this
case, for a QD spectrum a nonzero $\sin \theta$ can be generated by the
perturbations. If the neutrino masses are hierarchical, the value of $\sin
\theta$ induced by the corrections is negligible. The results for scenarios 1
and 2 are displayed in Figs.~\ref{fig:V13-V0-1} and \ref{fig:V13-V0-2},
respectively. For Dirac neutrinos, the value of $\sin \theta$ generated is
one order of magnitude smaller.

\begin{figure}[htb]
\begin{center}
\epsfig{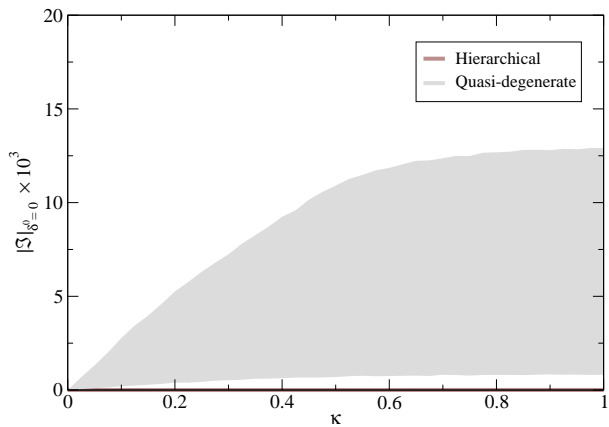}
\end{center}
\caption{Effect of the perturbations on $\mathcal{J}$,
for an initial vanishing Dirac phase, in
scenario 1.
\label{fig:ImQ1}}
\end{figure}

\begin{figure}[htb]
\begin{center}
\epsfig{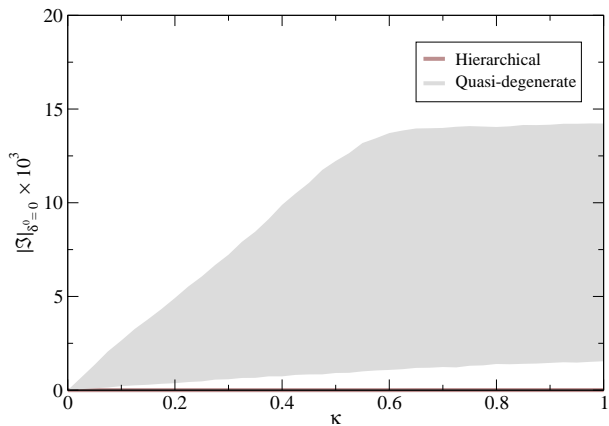}
\end{center}
\caption{Effect of the perturbations on $\mathcal{J}$,
for an initial vanishing Dirac phase, in
scenario 2.
\label{fig:ImQ2}}
\end{figure}

Finally, we consider the situation when $\delta = 0$ at the tree level but the
Majorana phases are not zero. In this limit, the $CP$ violating parameter
$\mathcal{J}$ in Eq.~(\ref{Jgen1}) vanishes. For a QD spectrum,
the corrections to the mass matrices induce a Dirac phase (provided at least one
of the Majorana phases is nonzero) large enough to yield
$\mathcal{J} \sim 10^{-2}$, which may
be observable by future long baseline neutrino oscillation
experiments \cite{Lindner:2002vt}. This is shown in
Figs.~\ref{fig:ImQ1} and \ref{fig:ImQ2}, and holds for $m_1 \gtrsim 0.7$ eV.
Setting one of the initial Majorana phases to zero does not eliminate this
effect, and for $\alpha=\pi/3$, $\beta=0$ the values of $\mathcal{J}$ obtained
are up to 0.02, even larger than the ones shown in Figs.~\ref{fig:ImQ1} and
\ref{fig:ImQ2}.
For a HI spectrum, the phase $\delta$ generated is negligible. In the case of
Dirac neutrinos, only the Dirac phase is physically meaningful, and the phases
of the random parameters $\lambda_i$, $\zeta_i$ and $\eta_i$ are not enough to
produce a relevant value of $\mathcal{J}$.

\section{Outlook}
\label{sec:5}

In this paper we have studied the possible effect of unknown
corrections from new physics on the neutrino mass hierarchy, mixing and $CP$
violation at low energies. We have focused on the case when neutrinos are
Majorana particles, but have discussed the results for Dirac neutrinos as well.
We have proposed a general parametrization of the
corrections to the tree-level mass matrices, based on weak basis invariance.
Using this parametrization, we have examined the consequences of adding
random perturbations to the mass matrices, as a means to explore the possible
effects that radiative corrections might yield.

We have analyzed the stability against corrections of the ratio of mass squared
differences and the mixing angles, for a hierarchical or quasi-degenerate
neutrino spectrum. We have found that these quantities are generally modified by
the perturbations, but the ratio
$\omega = \sin \theta/\sin \theta_\mathrm{atm}$ is remarkably stable even under
large corrections. This desirable property makes this quantity specially
suited for the experimental test of specific textures of neutrino mass matrices,
since it is hardly modified by unknown corrections to the tree-level matrices.
We have also examined the stability of some flavor relations predicted by
models in the literature.

The dependence of the deviations on the neutrino spectrum has also been
investigated. We have found that the region of neutrino masses $m \simeq 0.3$
eV is specially stable. For neutrino masses around this value, the possible
deviations in $r$, $\tan \theta_\odot$, $\tan
\theta_\mathrm{atm}$, $\sin \theta$ and the $CP$ violating phases are rather
small.
This mass region is of special interest, since it will be tested in upcoming
experiments.

We have addressed the question whether the tree-level mass matrices could
correspond to bi-maximal mixing, being the observed value
$\tan \theta_\odot \sim 0.6$ the result of radiative corrections. We have
demonstrated that, from an initial bi-maximal pattern, in the case of a QD
spectrum the corrections to the mass matrices can bring
down $\tan \theta_\odot$ to its experimental value while keeping
$\tan \theta_\mathrm{atm}$ very close to unity. This result holds both for
Majorana and Dirac neutrinos, and for a lightest neutrino with a mass larger
than $\simeq 0.6$ eV. Therefore, the possibility
of bi-maximal mixing cannot be excluded for a QD spectrum. On the other hand,
for a HI
spectrum $\tan \theta_\mathrm{atm}$ is modified but $\tan \theta_\odot$ not,
then bi-maximal mixing is highly unlikely in this case.

Other interesting limit examined is when $\sin \theta = 0$ at the tree level.
In this case, large corrections to the mass matrices could yield $\sin \theta
\sim 10^{-2}$ if the neutrinos are quasi-degenerate. On the contrary, we have
shown that for a HI spectrum or Dirac neutrinos the value of $\sin \theta$
generated by perturbations is negligible.

Finally, we have investigated the situation when the Dirac phase in the mixing
matrix vanishes at the tree level. In this case, the Majorana phases present
can induce a non-vanishing Dirac phase in the mixing matrix by means of the
perturbations. This Dirac phase is large enough to yield $\mathcal J \gtrsim
10^{-2}$, leading to observable $CP$ violation effects in long baseline
neutrino oscillation experiments.

\begin{acknowledgments}
This work has been supported by the European Community's Human Potential
Programme under contract HTRN--CT--2000--00149 Physics at Colliders and by FCT
through projects POCTI/FIS/36288/2000, POCTI/FNU/43793/2002 and CFIF--Plurianual
(2/91). The work of JAAS has been supported by FCT under grant
SFRH/BPD/12063/2003.
\end{acknowledgments}

\end{document}